\documentclass[12pt,hyperref,tightenlines,showpacs]{revtex4}%
\usepackage{amsfonts}
\usepackage[debug,pdftex]{hyperref}
\usepackage{amsmath}
\usepackage{amssymb}
\usepackage{graphicx}
\usepackage{bibmods}%
\providecommand{\U}[1]{\protect\rule{.1in}{.1in}}

\begin{document}
\title{On the Creation of Scalar Particles in a Flat Robertson-Walker Space-time}
\author{S. Haouat}
\email{s\_haouat@univ-jijel.dz \ }
\affiliation{\textit{LPTh, Department of Physics, University of Jijel, BP 98, Ouled Aissa,
Jijel 18000, Algeria.}}
\author{R. Chekireb}
\affiliation{\textit{LPTh, Department of Physics, University of Jijel, BP 98, Ouled Aissa,
Jijel 18000, Algeria.}}

\begin{abstract}
The problem of particle creation from vacuum in a flat Robertson-Walker
space-time is studied. Two sets of exact solutions for the Klein Gordon
equation are given when the scale factor is $a^{2}\left(  \eta\right)
=a+b\tanh\left(  \lambda\eta\right)  +c\tanh^{2}\left(  \lambda\eta\right)  $.
Then the canonical method based on Bogoliubov transformation is applied to
calculate the pair creation probability and the density number of created
particles. Some particular cosmological models such as radiation dominated
universe and Milne universe are discussed. For both cases the vacuum to vacuum
transition probability is calculated and the imaginary part of the effective
action is extracted.

\end{abstract}

\maketitle

\section{Introduction}

As we know there is no well defined theory of quantum gravity and the
quantization of gravitational fields is one of unsolved problems in
theoretical physics. Therefore the straightforward way to treat gravitational
effects is the quantum field theory in curved space-time which proved most
fruitful in describing interaction of matter with gravity and finding
physically meaningful quantities \cite{Birrel,fulling,Mukhanov,Grib,Parker}.
The quantum field theory in a curved space is the first approximation of
quantum gravity where the gravitational field described by the metric of the
space-time is treated as classical field and the matter is described by
quantum fields which propagate in such a gravitational field. It is well known
also that the most significant prediction of this theory is the phenomenon of
particle creation from the vacuum. Generally in curved space-time it is
difficult to define a physical vacuum state for the quantum field and when
this vacuum state is defined in the remote past it is usually unstable so that
it may differ from the vacuum state in the remote future, and spontaneous
creation of particles occurs .

The importance of particle production by gravitational fields comes from the
diversity of its applications in contemporary cosmology; Particle creation
effects could have consequences for early universe cosmology and could play a
role in the exit from inflationary universe and in the cosmic evolution
\cite{C1,C2,C3,C4,C5}. The application of this phenomenon in black hole
physics, such as Hawking radiation, is also well known \cite{Haw}.

In order to study the phenomenon of particle creation in gravitational fields
we found in literature several techniques such as the adiabatic approach
\cite{Parker1,Parker2,Parker3}, the Hamiltonian diagonalization method
\cite{Zeldo,Grib1,Grib2,Grib3,Haro1,Haro2}, the Green function approach
\cite{Gavrilov,Bukhbinder}, the Feynman path integral technique
\cite{Duru,Chitre}, the "in" and "out" states formalism
\cite{Garriga,Villalba1,Villalba2,Villalba3} as well as the semiclassical WKB
approximation \cite{Biswas1,Biswas2,Biswas3,Biswas4,Winitzki}. In addition we
cite the technique based on vacuum to vacuum transition amplitude and
Schwinger-like effective action \cite{Akhme,Kim}.

The aim of this paper is to study the creation of scalar particles in a flat
Robertson-Walker space-time provided with metric of the form
\begin{equation}
ds^{2}=dt^{2}-a^{2}\left(  t\right)  \left(  dx^{2}+dy^{2}+dz^{2}\right)
\label{1}%
\end{equation}
In terms of conformal time $\eta=\int dt/a\left(  t\right)  $ this metric
reads%
\begin{equation}
ds^{2}=C\left(  \eta\right)  \left[  d\eta^{2}-dx^{2}-dy^{2}-dz^{2}\right]
\label{2}%
\end{equation}
where the new scale factor $C\left(  \eta\right)  $ is defined from $a\left(
t\right)  $ as follows $C\left(  \eta\right)  =\tilde{a}^{2}\left(
\eta\right)  \equiv a^{2}\left[  t\left(  \eta\right)  \right]  .$ We choose
for the scale factor the form
\begin{equation}
C\left(  \eta\right)  =a+b\tanh\left(  \lambda\eta\right)  +c\tanh^{2}\left(
\lambda\eta\right)  \label{3}%
\end{equation}
where $a$, $b$ and $c$ are positive parameters.

We can see that this form is the generalization of various particular cases
found in literature; When $\allowbreak c=0,$ we have a cosmological model with
$a^{2}\left(  \eta\right)  =a+b\tanh\left(  \lambda\eta\right)  $ which has
been widely studied \cite{Bernard,Setare,Pascoal}. With a particular choice of
parameters $a$, $b$ and $c$ we get some models discussed in
\cite{Shapoor1,Shapoor2}. In addition, this universe becomes a radiation
dominated one when $a=b=0,$ $c=\frac{a_{0}^{4}}{4\lambda^{2}}$ and
$\lambda\rightarrow0$. We can also make connection with a Milne universe (i.e.
$a\left(  t\right)  =a_{1}t.$) when $c=0$, $\lambda=a_{1}$, $b=a=\frac
{a_{1}^{2}}{2\varepsilon}$ by making the change $\eta\rightarrow\eta+\frac
{\ln\varepsilon}{2\lambda}$and taking the limit $\varepsilon\longrightarrow0.$

In the first stage we introduce a scalar field propagating in Robertson-Walker
space-time and we give two sets of exact solutions for the Klein Gordon
equation. Next we use the relation between these two sets to determine the
probability of pair creation, the density number of created particles and the
vacuum persistence. Finally we discuss some particular examples.

\section{Scalar field and Klein Gordon equation}

Let us consider a scalar matter field $\Phi$ with mass $m$ subjected to the
gravitational field described by the metric $g_{\mu\nu}$. The dynamics of this
system is in general governed by the action%

\begin{equation}
S=\int d^{4}x\left[  \sqrt{-g}\left(  -g^{\mu\nu}\partial_{\mu}\Phi
\partial_{\nu}^{\ast}\Phi-m^{2}\Phi\Phi^{\ast}-V\left(  g\right)  \Phi
\Phi^{\ast}\right)  \right]  \label{4}%
\end{equation}
where $V\left(  g\right)  $ is a function of invariant combinations of the
metric tensor $g_{\mu\nu}$ and its partial derivatives \cite{Pavlov1,Pavlov}
\begin{equation}
V\left(  g\right)  =\xi_{1}R+\xi_{2}R_{GB}^{2}, \label{5}%
\end{equation}
where $R$ is the Ricci scalar, $\xi_{1}$ and $\xi_{2}$ are two numerical
factors and $R_{GB}^{2}$ is a scalar which explains the Gauss-Bonnet coupling%
\begin{equation}
R_{GB}^{2}=R^{2}-4R_{\mu\nu}R^{\mu\nu}+R_{\mu\nu\rho\sigma}R^{\mu\nu\rho
\sigma}. \label{6}%
\end{equation}
In such a case the Klein Gordon equation can be written in the form%
\begin{equation}
\frac{1}{\sqrt{-g}}\partial_{\mu}\left(  g^{\mu\nu}\sqrt{-g}\partial_{\nu}%
\Phi\right)  +\left(  m^{2}+V\left(  g\right)  \right)  \Phi=0. \label{7}%
\end{equation}
In the present work we interest only on conformally coupled scalar field (i.e.
$\xi_{1}=\frac{1}{6}$ and $\xi_{2}=0$). If we choose to work with conformal
time $\eta$ which is convenient to the present coupling and we introduce a new
field $\psi\left(  x\right)  $ so that%

\begin{equation}
\Phi\left(  x\right)  =\frac{1}{\sqrt{C\left(  \eta\right)  }}\psi\left(
x\right)  =\frac{1}{\sqrt{C\left(  \eta\right)  }}\chi\left(  \vec{x}\right)
\varphi\left(  \eta\right)  , \label{8}%
\end{equation}
where $\chi\left(  \vec{x}\right)  $ has, in the case of flat space-time, the
form of a plane wave $\chi\left(  \vec{x}\right)  \sim\exp\left(  i\vec
{k}.\vec{r}\right)  ,$ we can obtain the simplified equation
\begin{equation}
\left[  \frac{d^{2}}{d\eta^{2}}+\omega^{2}(\eta)\right]  \varphi\left(
\eta\right)  =0, \label{9}%
\end{equation}
with%
\begin{equation}
\omega^{2}(\eta)=k^{2}+m^{2}C(\eta). \label{10}%
\end{equation}
Since equation (\ref{9}) is of second order there are only two independent
solutions and all other solutions can be expressed in terms of these two
independent ones. Here we want to find two sets of independent solutions so
that the two functions $\varphi_{in}^{\pm}$ of the first set behave like
positive and negative energy states at $\eta\rightarrow-\infty$ and the two
functions $\varphi_{out}^{\pm}$ of the second set behave like positive and
negative energy states at $\eta\rightarrow+\infty$. Therefore, before to look
for exact solutions for equation (\ref{9}), let us first, examine their
behavior when $\eta\rightarrow\pm\infty.$ We easily obtain
\begin{align}
\varphi_{in}^{\epsilon}\left(  \eta\right)   &  =\exp\left(  -i\epsilon
\omega_{in}\eta\right) \label{11}\\
\varphi_{out}^{\epsilon}\left(  \eta\right)   &  =\exp\left(  -i\epsilon
\omega_{out}\eta\right)  , \label{12}%
\end{align}
where $\epsilon$ indicates the positive or the the negative frequency mode (
$\epsilon=\pm1$) and $\omega_{out}$ and $\omega_{in}$ are given by%
\begin{align}
\omega_{in}  &  =\sqrt{k^{2}+m^{2}\left(  a+c-b\right)  }\label{13}\\
\omega_{out}  &  =\sqrt{k^{2}+m^{2}\left(  a+c+b\right)  }. \label{14}%
\end{align}
Note that behaviors in (\ref{11}) and (\ref{12}) are the same as
semi-classical solutions obtained from Hamilton-Jacobi equation because the
space-time is asymtotically Minkowskian.

Now in order to solve equation (\ref{9}) we make the change $\eta
\rightarrow\xi,$ where%

\begin{equation}
\xi=\frac{1+\tanh\left(  \lambda\eta\right)  }{2}. \label{15}%
\end{equation}
The resulting equation that takes the form%

\begin{align}
&  \left[  \frac{\partial^{2}}{\partial\xi^{2}}+\left(  \frac{1}{\xi}-\frac
{1}{1-\xi}\right)  \frac{\partial}{\partial\xi}+\left(  \frac{\omega_{in}^{2}%
}{4\lambda^{2}}\frac{1}{\xi}-\frac{m^{2}c}{\lambda^{2}}\right.  \right.
\nonumber\\
&  \left.  \left.  \left.  +\frac{\omega_{out}^{2}}{4\lambda^{2}}\frac
{1}{\left(  1-\xi\right)  }\right)  \frac{1}{\xi\left(  1-\xi\right)
}\right]  \tilde{\varphi}\left(  \xi\right)  =0\right.  \label{16}%
\end{align}
is a Riemann type equation \cite{Grad}
\begin{align}
&  \left[  \frac{\partial^{2}}{\partial\xi^{2}}+\left(  \frac{1-\alpha
_{1}-\alpha_{1}^{\prime}}{\xi}-\frac{1-\alpha_{3}-\alpha_{3}^{\prime}}{1-\xi
}\right)  \frac{\partial}{\partial\xi}+\right. \nonumber\\
&  \left.  \left.  \left(  \frac{\alpha_{1}\alpha_{1}^{\prime}}{\xi}%
-\alpha_{2}\alpha_{2}^{\prime}+\frac{\alpha_{3}\alpha_{3}^{\prime}}{1-\xi
}\right)  \frac{1}{\xi\left(  1-\xi\right)  }\right]  \tilde{\varphi}\left(
\xi\right)  =0\right.  \label{17}%
\end{align}
where%

\begin{equation}%
\begin{array}
[c]{l}%
\alpha_{1}=-\alpha_{1}^{\prime}=i\frac{\omega_{in}}{2\lambda}\\
\alpha_{3}=-\alpha_{3}^{\prime}=i\frac{\omega_{in}}{2\lambda}\\
\alpha_{2}=1-\alpha_{2}^{\prime}=\frac{1}{2}+\sqrt{\frac{m^{2}c}{\lambda^{2}%
}-\frac{1}{4}},
\end{array}
\label{18}%
\end{equation}
with the condition $\alpha_{1}+\alpha_{1}^{\prime}+\alpha_{2}+\alpha
_{2}^{\prime}+\alpha_{3}+\alpha_{3}^{\prime}=1.$

Following \cite{Grad} we can find for equation (\ref{9}) several sets of
solutions that can be written in terms of hypergeometric functions. Taking
into account the behavior of positive and negative energy states we can
classify our to sets as follows; for the "in" states we have%

\begin{align}
\tilde{\varphi}_{in}^{+}\left(  \xi\right)   &  =\frac{1}{\sqrt{2\omega_{in}}%
}\xi^{-i\frac{\omega_{in}}{2\lambda}}(1-\xi)^{i\frac{\omega_{out}}{2\lambda}%
}\nonumber\\
&  F\left(  \frac{1}{2}+i\frac{\omega_{-}}{\lambda}+i\delta,\frac{1}{2}%
+i\frac{\omega_{-}}{\lambda}-i\delta;1-i\frac{\omega_{in}}{\lambda}%
;\xi\right)  \label{19}%
\end{align}
and%
\begin{align}
\tilde{\varphi}_{in}^{-}\left(  \xi\right)   &  =\frac{1}{\sqrt{2\omega_{in}}%
}\xi^{i\frac{\omega_{in}}{2\lambda}}(1-\xi)^{-i\frac{\omega_{out}}{2\lambda}%
}\nonumber\\
&  F\left(  \frac{1}{2}-i\frac{\omega_{-}}{\lambda}+i\delta,\frac{1}{2}%
-i\frac{\omega_{-}}{\lambda}-i\delta;1+i\frac{\omega_{in}}{\lambda}%
;\xi\right)  , \label{20}%
\end{align}
with%
\begin{equation}
\omega_{\pm}=\frac{\omega_{out}\pm\omega_{in}}{2} \label{21}%
\end{equation}
and%
\begin{equation}
\delta=\frac{1}{2}\sqrt{\frac{4m^{2}c}{\lambda^{2}}-1}. \label{22}%
\end{equation}
The factors $\left(  2\omega_{in}\right)  ^{-1/2}$ and $\left(  2\omega
_{out}\right)  ^{-1/2}$ are determined by the use of the following
normalization condition
\begin{equation}
\varphi_{k}^{\ast}\dot{\varphi}_{k}-\varphi_{k}\dot{\varphi}_{k}^{\ast}=2i,
\label{23}%
\end{equation}
which explains the conservation of the Klein Gordon particle current.

For the "out" states we have%

\begin{align}
\tilde{\varphi}_{out}^{+}\left(  \xi\right)   &  =\frac{1}{\sqrt{2\omega
_{out}}}\xi^{-i\frac{\omega_{in}}{2\lambda}}(1-\xi)^{i\frac{\omega_{out}%
}{2\lambda}}\nonumber\\
&  F\left(  \frac{1}{2}+i\frac{\omega_{-}}{\lambda}+i\delta,\frac{1}{2}%
+i\frac{\omega_{-}}{\lambda}-i\delta;1+i\frac{\omega_{out}}{\lambda}%
;1-\xi\right)  \label{24}%
\end{align}
and%
\begin{align}
\tilde{\varphi}_{out}^{-}\left(  \xi\right)   &  =\frac{1}{\sqrt{2\omega
_{out}}}\xi^{i\frac{\omega_{in}}{2\lambda}}(1-\xi)^{-i\frac{\omega_{out}%
}{2\lambda}}\nonumber\\
&  F\left(  \frac{1}{2}-i\frac{\omega_{-}}{\lambda}+i\delta,\frac{1}{2}%
-i\frac{\omega_{-}}{\lambda}-i\delta;1-i\frac{\omega_{out}}{\lambda}%
;1-\xi\right)  . \label{25}%
\end{align}

Having shown how to solve Klein Gordon equation for spinless particle
conformally coupled to gravitational field described by the metric (\ref{2}),
let us investigate the particle creation process. In the next section we
quantify the scalar matter field and we use the solutions of the field
equation to analyze the problem in question.

\section{Field quantization and particle creation}

In the first stage we write the field operator in it's Fourier decomposition
\begin{equation}
\hat{\psi}(\vec{x},\eta)=\frac{1}{\sqrt{2}}\int d^{3}k\left[  \varphi
_{k}^{\ast}\left(  \eta\right)  \chi_{_{k}}\left(  \vec{x}\right)  \hat{a}%
_{k}+\varphi_{k}\left(  \eta\right)  \chi_{_{k}}^{\ast}\left(  \vec{x}\right)
\hat{b}_{k}^{+}\right]  \label{26}%
\end{equation}
where, in canonical quantization formalism, the operators $\hat{a}_{k}$ and
$\hat{b}_{k}$ verify the following commutation relation%

\begin{equation}
\left[  \hat{a}_{k},\hat{a}_{k^{\prime}}^{+}\right]  =\left[  \hat{b}_{k}%
,\hat{b}_{k^{\prime}}^{+}\right]  =\delta\left(  \vec{k}-\vec{k}^{\prime
}\right)  . \label{27}%
\end{equation}
With the help of the normalization condition (\ref{23}) we can find without
difficulties the following expression of the Hamiltonian associated with the
scalar field system%
\begin{equation}
H=\frac{1}{2}\int d^{3}k\left[  E_{k}\left(  \eta\right)  \left(  \hat{a}%
_{k}\hat{a}_{k}^{+}+\hat{b}_{k}^{+}\hat{b}_{k}\right)  +F_{k}^{^{\ast}}\left(
\eta\right)  \hat{b}_{k}\hat{a}_{k}+F_{k}\left(  \eta\right)  \hat{a}_{k}%
^{+}\hat{b}_{k}^{+}\right]  \label{28}%
\end{equation}
with%
\begin{align}
E_{k}\left(  \eta\right)   &  =\left\vert \dot{\varphi}_{k}\left(
\eta\right)  \right\vert ^{2}+\omega_{k}^{2}\left(  \eta\right)  \left\vert
\varphi_{k}\left(  \eta\right)  \right\vert ^{2}\label{29}\\
F_{k}\left(  \eta\right)   &  =\dot{\varphi}_{k}^{2}\left(  \eta\right)
+\omega_{k}^{2}\left(  \eta\right)  \varphi_{k}^{2}\left(  \eta\right)  .
\label{30}%
\end{align}
Here we remark that $H$ is not diagonal at any time. For $\eta\rightarrow
\pm\infty$, however, $F_{k}\left(  \eta\right)  =0$ and $H$ \ becomes
diagonal. In this situation we can define two vacuum states $\left\vert
0_{in}\right\rangle $ and $\left\vert 0_{out}\right\rangle .$ The state
$\left\vert 0_{in}\right\rangle $ is an initial quantum vacuum state in the
remote past with respect to a static observer and $\left\vert 0_{out}%
\right\rangle $ is a final quantum vacuum state in the remote future with
respect to the same observer. This gives some vacuum instability which leads
to particle creation.

Now, as we mentioned above, in order to determine the probability of pair
creation and the density of created particles we use the solutions of
the\ Klein Gordon equation by considering the canonical method based the
relation between "in" and "out" states. In other wards Bogoliubov
transformation connecting the "in" with the "out" states can be projected in
Fock space to be converted into relation between the creation and annihilation
operators; therefore, the probability of pair creation and the density of
created particles will be given in terms of Bogoliubov coefficients.

So to get connection between "in" and "out" modes let us use the relation
between hypergeometric functions \cite{Grad}%
\begin{align}
F\left(  u,v;w;\xi\right)   &  =\frac{\Gamma\left(  w\right)  \Gamma\left(
w-v-u\right)  }{\Gamma\left(  w-u\right)  \Gamma\left(  w-v\right)  }F\left(
u,v;u+v-w+1;1-\xi\right)  ~~\nonumber\\
&  +\left(  1-\xi\right)  ^{w-u-v}\frac{\Gamma\left(  \gamma\right)
\Gamma\left(  u+v-w\right)  }{\Gamma\left(  u\right)  \Gamma\left(  v\right)
}\nonumber\\
&  F\left(  w-u,w-v;w-v-u+1;1-\xi\right)  \label{31}%
\end{align}
and the property%
\begin{equation}
F\left(  u,v;w;\xi\right)  =\left(  1-\xi\right)  ^{w-u-v}F\left(
w-u,w-v;w;\xi\right)  \label{32}%
\end{equation}
to obtain the so-called Bogoliubov transformation connecting "in" and "out" states%

\begin{align}
\varphi_{in}^{+}  &  =\alpha\varphi_{out}^{+}+\beta\varphi_{out}^{-}%
\label{33}\\
\varphi_{in}^{-}  &  =\beta^{\ast}\varphi_{out}^{+}+\alpha^{\ast}\varphi
_{out}^{-}, \label{34}%
\end{align}
where the Bogoliubov coefficients $\alpha$ and $\beta$ are given by%
\begin{equation}
\alpha=\sqrt{\frac{\omega_{out}}{\omega_{in}}}\frac{\Gamma\left(
1-i\frac{\omega_{in}}{\lambda}\right)  \Gamma\left(  -i\frac{\omega_{out}%
}{\lambda}\right)  }{\Gamma\left(  \frac{1}{2}-i\frac{\omega_{+}}{\lambda
}-i\frac{1}{2}\sqrt{\frac{4m^{2}c}{\lambda^{2}}-1}\right)  \Gamma\left(
\frac{1}{2}-i\frac{\omega_{+}}{\lambda}+i\frac{1}{2}\sqrt{\frac{4m^{2}%
c}{\lambda^{2}}-1}\right)  } \label{35}%
\end{equation}
and%
\begin{equation}
\beta=\sqrt{\frac{\omega_{out}}{\omega_{in}}}\frac{\Gamma\left(
1-i\frac{\omega_{in}}{\lambda}\right)  \Gamma\left(  i\frac{\omega_{out}%
}{\lambda}\right)  }{\Gamma\left(  \frac{1}{2}+i\frac{\omega_{-}}{\lambda
}-i\frac{1}{2}\sqrt{\frac{4m^{2}c}{\lambda^{2}}-1}\right)  \Gamma\left(
\frac{1}{2}+i\frac{\omega_{-}}{\lambda}+i\frac{1}{2}\sqrt{\frac{4m^{2}%
c}{\lambda^{2}}-1}\right)  } \label{36}%
\end{equation}
with $\left\vert \alpha\right\vert ^{2}-\left\vert \beta\right\vert ^{2}=1.
$The relation between the creation and annihilation operators is then%

\begin{align}
a_{out}  &  =\alpha\ a_{in}+\beta^{\ast}b_{in}^{+}\label{37}\\
b_{out}^{+}  &  =\beta\ a_{in}+\alpha^{\ast}b_{in}^{+}. \label{38}%
\end{align}
For the process of particle creation the probability amplitude that we want to
calculate is defined by%

\begin{equation}
\mathcal{A}=\left\langle 0_{out}\left\vert a_{out}b_{out}\right\vert
0_{in}\right\rangle . \label{39}%
\end{equation}
Taking into account that\qquad%

\begin{equation}
b_{out}=\frac{1}{\alpha^{\ast}}b_{in}+\frac{\beta^{\ast}}{\alpha^{\ast}%
}a_{out}^{+} \label{40}%
\end{equation}
we obtain%

\begin{equation}
\mathcal{A}=\left\langle 0_{out}\left\vert a_{out}b_{out}\right\vert
0_{in}\right\rangle =\frac{\beta^{\ast}}{\alpha^{\ast}}\left\langle
0_{out}\mid0_{in}\right\rangle . \label{41}%
\end{equation}
The probability to create one pair of particles from vacuum is then%

\begin{equation}
\mathcal{P}_{creat.}\left(  k\right)  =\left\vert \frac{\beta^{\ast}}%
{\alpha^{\ast}}\right\vert ^{2}. \label{42}%
\end{equation}
Using the following properties of the Gamma functions \cite{Grad}%
\begin{equation}
\Gamma(z+1)=z\Gamma(z), \label{43}%
\end{equation}%
\begin{equation}
\left\vert \Gamma(ix)\right\vert ^{2}=\frac{\pi}{x\sinh\pi x} \label{44}%
\end{equation}
and%
\begin{equation}
\left\vert \Gamma(\frac{1}{2}+ix)\right\vert ^{2}=\frac{\pi}{\cosh\pi x}
\label{45}%
\end{equation}
we arrive at
\begin{equation}
\mathcal{P}_{creat.}\left(  k\right)  =\frac{\cosh\left(  2\pi\frac{\omega
-}{\lambda}\right)  +\cosh\left(  2\pi\delta\right)  }{\cosh\left(  2\pi
\frac{\omega+}{\lambda}\right)  +\cosh\left(  2\pi\delta\right)  }. \label{46}%
\end{equation}
Let $\mathcal{C}_{k}$ to be the probability to have no pair creation in the
state $k.$The quantity $\mathcal{C}_{k}\left(  P_{creat.}\right)  ^{n}$ is
then the probability to have only $n$ pairs in the state $k.$ We have%

\begin{equation}
\mathcal{C}_{k}+\mathcal{C}_{k}P_{creat.}+\mathcal{C}_{k}\left(
P_{creat.}\right)  ^{2}+\mathcal{C}_{k}\left(  P_{creat.}\right)  ^{3}+...=1
\label{47}%
\end{equation}
or simply%

\begin{equation}
\mathcal{C}_{k}=1-P_{creat.}. \label{48}%
\end{equation}
Being aware of $\left\vert \frac{\beta^{\ast}}{\alpha^{\ast}}\right\vert
^{2}+\left\vert \frac{1}{\alpha^{\ast}}\right\vert ^{2}=1,$ we can find the
vacuum persistence which reads%
\begin{align*}
\mathcal{C}_{k}  &  =\left\vert \frac{1}{\alpha^{\ast}}\right\vert ^{2}\\
&  =\frac{\cosh\left(  2\pi\frac{\omega+}{\lambda}\right)  -\cosh\left(
2\pi\frac{\omega-}{\lambda}\right)  }{\cosh\left(  2\pi\frac{\omega+}{\lambda
}\right)  +\cosh\left(  2\pi\delta\right)  }.
\end{align*}
Another important result is the density number of created particles%
\begin{equation}
n\left(  k\right)  =\left\langle 0_{in}\left\vert a_{out}^{+}a_{out}%
\right\vert 0_{in}\right\rangle =\left\vert \beta\right\vert ^{2} \label{50}%
\end{equation}
A simple calculation gives%

\begin{equation}
n\left(  k\right)  =\frac{\cosh\left(  2\pi\frac{\omega_{-}}{\lambda}\right)
+\cosh\left(  \pi\sqrt{\frac{4m^{2}c}{\lambda^{2}}-1}\right)  }{\cosh\left(
2\pi\frac{\omega+}{\lambda}\right)  -\cosh\left(  2\pi\frac{\omega-}{\lambda
}\right)  }. \label{51}%
\end{equation}
Here we note that the density number of created particles can be written as%
\begin{equation}
n\left(  k\right)  =\frac{1}{\left\vert \frac{\alpha}{\beta}\right\vert
^{2}-1}, \label{52}%
\end{equation}
and for large frequencies $n\left(  k\right)  $ becomes a thermal
Bose-Einstein distribution%
\begin{equation}
n\left(  k\right)  =\frac{1}{\exp\left(  \frac{2\pi}{\lambda}\omega
_{in}\right)  -1}. \label{53}%
\end{equation}
From equation (\ref{51}) one can see that the pair creation process is more
important for low impulsion and $n\left(  k\right)  \rightarrow0$ when
$k\rightarrow\infty$ (i.e. $k>>1$). Also, for $a-b+c\neq0$ and $a+b>c,$ it is
clear that $n\left(  k\right)  \rightarrow0$ when $m>>1$.

\section{Special cases}

In order to illustrate our calculation let us in this section discuss some
particular cosmological models such as radiation dominated universe and the
Milne universe which are of interest and have been studied by other authors
and by using other techniques. We start by radiation dominated universe.

\subsection{Radiation dominated universe}

Considering the case when $a=b=0$ and $c=\frac{a_{0}^{4}}{4\lambda^{2}}$ and
taking the limit $\lambda\rightarrow0,$ we make connection with the well-known
phase of radiation dominated universe%
\begin{equation}
C\left(  \eta\right)  =\frac{a_{0}^{4}}{4}\eta^{2} \label{54}%
\end{equation}
and%
\begin{equation}
a\left(  t\right)  =a_{0}\sqrt{t}. \label{55}%
\end{equation}
In this case the probability to create a pair of particles will be%

\begin{equation}
\mathcal{P}_{rad.}\left(  k\right)  =\lim_{\lambda\rightarrow0}\frac
{\exp\left(  \pi\sqrt{\frac{4m^{2}c}{\lambda^{2}}-1}-\frac{2\pi\omega}%
{\lambda}\right)  }{1+\exp\left(  \pi\sqrt{\frac{4m^{2}c}{\lambda^{2}}%
-1}-\frac{2\pi\omega}{\lambda}\right)  }. \label{56}%
\end{equation}
Taking into account that%
\begin{equation}
\pi\sqrt{\frac{4m^{2}c}{\lambda^{2}}-1}-\frac{2\pi\omega}{\lambda}%
=m\frac{a_{0}^{2}}{2\lambda^{2}}\sqrt{1+\frac{4\lambda^{2}k^{2}}{m^{2}%
a_{0}^{4}}}-\frac{1}{2}m\frac{a_{0}^{2}}{\lambda^{2}}\sqrt{1-\frac{\lambda
^{4}}{a_{0}^{4}m^{2}}}\approx\frac{k^{2}}{ma_{0}^{2}} \label{57}%
\end{equation}
we get%
\begin{equation}
\mathcal{P}_{rad.}\left(  k\right)  =\frac{\exp\left[  -\pi\frac{2k^{2}%
}{ma_{0}^{2}}\right]  }{1+\exp\left[  -\pi\frac{2k^{2}}{ma_{0}^{2}}\right]  }.
\label{58}%
\end{equation}
The density number of created particles will be
\begin{equation}
n\left(  k\right)  =\exp\left(  -\frac{2\pi k^{2}}{ma_{_{0}}^{2}}\right)  .
\label{59}%
\end{equation}
This results coincide exactly with those found in literature \cite{Duru}.

For the vacuum to vacuum probability we can write%

\begin{align}
\mathcal{P}_{vac\rightarrow vac} &  =\exp\left(  -2\operatorname{Im}%
S_{eff}\right)  \nonumber\\
&  =\prod_{k}\mathcal{C}_{k}\nonumber\\
&  =\prod_{k}\exp\left[  -\ln\left(  1+\sigma\right)  \right]  \nonumber\\
&  =\exp\left[  -\sum_{k}\ln\left(  1+\sigma\right)  \right]  ,\label{60}%
\end{align}
where $\sigma=\exp\left(  -\frac{2\pi k^{2}}{ma_{_{0}}^{2}}\right)  .$
Consequently, we have%

\begin{equation}
2\operatorname{Im}S_{eff}=\sum_{k}\ln\left(  1+\sigma\right)  . \label{61}%
\end{equation}
Expanding the quantity $\ln(1+\sigma),$ we get%

\begin{equation}
2\operatorname{Im}S_{eff}=\int\frac{d^{3}k}{\left(  2\pi\right)  ^{3}}%
\ \sum_{n=1}\frac{\left(  -1\right)  ^{n+1}}{n}\exp\left(  -n\pi\frac{2k^{2}%
}{a_{0}^{2}m}\right)  . \label{62}%
\end{equation}
By doing integration over momentum $k$%

\begin{equation}
2\operatorname{Im}S_{eff}=\left(  \frac{a_{0}^{2}m}{8\pi^{2}}\right)
^{\frac{3}{2}}\sum_{n=1}\frac{\left(  -1\right)  ^{n+1}}{n^{\frac{5}{2}}}
\label{63}%
\end{equation}
and summing over $n,$ we obtain the following result%

\begin{equation}
2\operatorname{Im}S_{eff}=0.867\left(  \frac{ma_{0}^{2}}{8\pi^{2}}\right)
^{\frac{3}{2}}. \label{64}%
\end{equation}
It is clear that the effect becomes important as soon as the factor $a_{0}$
approaches the critical value%
\begin{equation}
a_{cr}^{2}\sim\frac{8\pi^{2}}{m} \label{65}%
\end{equation}
and the pair production rate is of order of $\left(  \frac{a_{0}}{a_{cr}%
}\right)  ^{3}.$

\subsection{Milne universe}

Now we consider the example of Milne universe defined by the scale factor
$a\left(  t\right)  =a_{1}t.$ In term of conformal time this scale factor
reads%
\begin{equation}
C\left(  \eta\right)  =a^{2}\left(  \eta\right)  =a_{1}^{2}\exp\left(
2a_{1}\eta\right)  . \label{66}%
\end{equation}
As is mentioned above this case may be obtained from (\ref{3}) by considering
the case $c=0$, $\lambda=a_{1}$, $b=a=\frac{a_{1}^{2}}{2\varepsilon}$, making
the change $\eta\rightarrow\eta+\frac{\ln\varepsilon}{2\lambda}$and taking the
limit $\varepsilon\longrightarrow0.$ In these conditions the probability to
have one pair creation will be
\begin{equation}
\mathcal{P}_{creat.}=\exp\left(  -\frac{2\pi}{a_{1}}k\right)  , \label{67}%
\end{equation}
which is in agreement with the same probability calculated by other authors
\cite{Duru}.

The vacuum persistence is%
\begin{equation}
\mathcal{C}_{k}=1-\exp\left(  -\frac{2\pi}{a_{1}}k\right)  \label{68}%
\end{equation}
and the vacuum to vacuum probability takes the form%

\begin{align}
\exp\left(  -2\operatorname{Im}S_{eff}\right)   &  =\prod_{k}\mathcal{C}%
_{k}\nonumber\\
&  =\exp\left[  \sum_{k}\ln\left(  1-\exp\left(  -\frac{2\pi}{a_{1}}k\right)
\right)  \right]  \label{69}%
\end{align}
from which we draw the imaginary part of the effective action%
\begin{equation}
2\operatorname{Im}S_{eff}=-\sum_{k}\ln\left(  1-\exp\left(  -\frac{2\pi}%
{a_{1}}k\right)  \right)  \label{70}%
\end{equation}
Making the Taylor expansion of the logarithm function%
\begin{equation}
2\operatorname{Im}S_{eff}=\int\frac{d^{3}k}{\left(  2\pi\right)  ^{3}}%
\ \sum_{n=1}\frac{1}{n}\exp\left(  -n\frac{2\pi}{a_{1}}k\right)  \label{71}%
\end{equation}
and summing over $n$ we obtain%
\begin{equation}
2\operatorname{Im}S_{eff}=1.2\times\frac{a_{1}^{2}}{4\pi^{4}}. \label{72}%
\end{equation}
Here we remark that the pair production rate is of order of $a_{1}^{2}.$

\subsection{Other particular cases}

Since the creation of superheavy particles with the mass of the Grand
Unification scale in the early Universe has many important cosmological
consequences \cite{grpa1,grpa2,grpa3} we want to discuss in this paragraph
cases with $m>>1.$ In general when $m>>1,$ we have shown that $n\left(
k\right)  \rightarrow0$. However, when $a-b+c=0$ or $c>a+b$ we remark that
$n\left(  k\right)  \rightarrow C^{st}>0$ even if $m>>1.$ This implies that it
is possible to create superheavy particles in this model.

Let us mention that if we put $c=0,$ we obtain the same results as those found
in \cite{Setare,Pascoal,Shapoor1,Shapoor2} and when $a=\left(  \frac
{1+\varepsilon}{2}\right)  ^{2}$; $b=\frac{1-\varepsilon^{2}}{2}$; $c=\left(
\frac{1-\varepsilon}{2}\right)  ^{2}$\ our results coincide with those of
\cite{Shapoor1,Shapoor2}.

\section{Conclusion}

In this paper we have studied the creation of scalar particles in some flat
Robertson-Walker space-times. We have considered the canonical method based on
Bogoliubov transformation connecting the "in" with the "out" states.
$\allowbreak$We have given two sets of exact solutions for the Klein Gordon
field equation and we have used these solutions to expressed the probability
of pair creation and the density of created particles in some closed forms.
Then we have discussed some particular cosmological models where our results
become the same as those obtained by other authors. For the radiation
dominated universe and the Milne universe we have calculated the vacuum to
vacuum transition probability and we have extracted the nonvanishing imaginary
term of the effective action that means that created particles are real and
not virtual ones. We have shown also that creation of superheavy particles may
be important in the case when $a-b+c=0$ or $c>a+b$.

\end{document}